\documentclass[sigconf]{acmart}

\usepackage{booktabs} 
\usepackage{color, algorithm, algorithmic}
\usepackage{amsmath, xspace}
\usepackage{algorithmic, algorithm}

\usepackage{natbib}

\newcommand{\JDAM}{\textrm{\small{JDAM}}\xspace}
\newcommand{\DDS}{\textrm{DDS}}
\newcommand{\target}[1]{#1^\odot}
\newcommand{\ie}{i.e.,}
\setcopyright{none}

\newcommand{\WHILE}{\textbf{while }}
\newcommand{\FOR}{\textbf{for }}

\newcommand{\IF}{\textbf{if }}

\newcommand{\SP}{${}$\hspace{0.5cm}}

\begin{document}
\title{Construction of Directed 2K Graphs}

\author{B\'alint Tillman}
\affiliation{
  \institution{University of California, Irvine}
}
\email{tillmanb@uci.edu}

\author{Athina Markopoulou}
\affiliation{
  \institution{University of California, Irvine}
}
\email{athina@uci.edu}

\author{Carter T. Butts}
\affiliation{
  \institution{University of California, Irvine}
  }
\email{buttsc@uci.edu}

\author{Minas Gjoka}
\affiliation{
  \institution{Google}
  }
\email{minasgjoka@google.com}

\renewcommand{\shortauthors}{B. Tillman et al.}

\settopmatter{printacmref=false}
\renewcommand\footnotetextcopyrightpermission[1]{}
\fancyhead[LE]{\shorttitle}
\fancyhead[RO]{\shortauthors}

\begin{abstract}
We study the problem of constructing synthetic graphs that resemble real-world directed graphs in terms of their degree correlations. We define  the problem of {\em directed 2K construction (D2K)} that takes as input the directed degree sequence (DDS) and a joint degree and attribute matrix (JDAM) so as to capture degree correlation specifically in directed graphs. We provide necessary and sufficient conditions to decide whether a target D2K is realizable, and we design an efficient algorithm that creates realizations with that target D2K. We evaluate our algorithm in creating synthetic graphs that target real-world directed graphs (such as Twitter) and we show that it brings significant benefits compared to state-of-the-art approaches.
\end{abstract}

%
%

\ccsdesc[500]{Mathematics of computing~Graph algorithms}

\keywords{Directed graphs, graph realizations, construction algorithms}

\maketitle

\section{Introduction}

It is often desirable to generate synthetic graphs that resemble real-world networks w.r.t. certain properties of interest. For example, researchers often want to simulate a process on a realistic network topology, and they may not have access to a real-world network, or they may want to generate several different realizations of the graphs of interest. In this paper, we focus specifically on directed graphs that appear in many application scenarios including, but not limited to, online social networks such as Twitter ({\em e.g.,} referring to the follower relations or to actual communication among users).  

There is a large body of work, both classic (\cite{erdHos1960grafok},\cite{havel1955poznamka},\cite{hakimi1962realizability},\cite{taylor1980constrained}) and recent (dK-series \cite{mahadevan2006systematic},\cite{orsini2015quantifying}, PAM \cite{erdHos2015graph})  that has studied the problem of constructing realizations of undirected graphs that exhibit exactly some target structural properties such as a given degree distribution or a given joint degree matrix. In this paper, we adopt the dK-series framework  \cite{mahadevan2006systematic},\cite{orsini2015quantifying}, which provides an elegant way to trade accuracy (in terms of graph properties) for complexity (in generating graph realizations). Construction of dK-graphs is well understood ({\em i.e.,} efficient algorithms and realizability conditions are known) for 1K (graphs with a given degree distribution) and 2K ( graphs with a given joint degree matrix). For $d>2$ (which is necessary for capturing the clustering exhibited in social networks), we recently proved that the problem is NP-hard \cite{devanny2016computational} but we also developed efficient heuristics \cite{gjoka2015construction}. In contrast, construction is not well-understood for directed graphs: results are known for construction  of graphs with a target directed degree sequence \cite{gale1957theorem}, \cite{fulkerson1960zero}, but there is no notion of directed degree correlation or directed dK-series for $d\geq2$.  

In this paper, we address this problem. We define two notions of degree correlation in directed 2K graphs, namely directed 2K (D2K), and its special case D2Km. D2K includes the notion of directed degree sequence and builds on an old trick (mapping directed graphs to bipartite undirected graphs) to also express degree-correlation via a joint degree-attribute matrix (JDAM) for the bipartite graph. This problem definition lends itself naturally to techniques we previously developed for undirected 2K \cite{gjoka2015construction}, an observation that we exploit to develop (i) necessary and sufficient realizability conditions  and (ii) an efficient algorithm that constructs D2K realizations. Our D2K approach advances the state-of-the-art in modeling and simulating realistic directed graphs, especially in the context of online social networks.

The outline of the rest of the paper is as follows. Section \ref{sec:related} summarizes related work. Section \ref{sec:problem} defines the Directed 2K problem (D2K and its special case D2Km). Section \ref{sec:algorithm} provides realizability conditions for D2K and an efficient algorithm for constructing such realizations. Section \ref{sec:results} applies the algorithm to construct Directed 2K graphs that resemble real world graphs, and demonstrates the advantages of our approach compared to state-of-the-art approaches. Section \ref{sec:conclusion} concludes the paper.

\section{Related work}\label{sec:related}

We adopt the systematic framework of dK-series \cite{mahadevan2006systematic}, which characterizes the properties of a graph using a series of probability distributions specifying all degree correlations within d-sized, simple, and connected subgraphs of a given graph G. In this framework, higher values of $d$ capture progressively more properties of G at the cost of more complex representation of the probability distribution. The dK-series exhibit two desired properties: inclusion (a dK distribution includes all properties defined by any $d'$K distribution, $\forall d' < d$ and convergence ($n$K, where $n = |V |$ specifies the entire graph, within isomorphism). 

We focus on graph construction approaches that produce simple graphs with prescribed target distributions, unlike the stochastic approach presented by \cite{dorogovtsev2003networks} or the configuration model in \cite{aiello2000random}. Algorithms of known time complexity exist for $d \leq 2$, and Monte Carlo Markov Chain (MCMC) approaches are used for $d > 2$.

{\bf 0K Construction.} 0K describes graphs with prescribed number of nodes and edges. This notion translates to simple Erd{\H{o}}s-R{\'e}nyi (ER) graphs with fixed number of edges. There is a simple extension of ER graphs to generate not just undirected but directed graphs as well, which we will use in our evaluation.

{\bf 1K Construction.} Degree sequences are equivalent to 1K as defined in the dK-series. Because degree sequences have been studied since the 1950s, we only focus on the most relevant results.  The realizability conditions for degree sequences were given by the Erd{\H{o}}s-Gallai theorem \cite{erdHos1960grafok}, and first algorithm to produce a single realization by Havel-Hakimi \cite{havel1955poznamka},\cite{hakimi1962realizability}. More recently, importance sampling algorithms were proposed in \cite{blitzstein2011sequential} and \cite{del2010efficient}.

{\bf 2K Construction.} Joint Degree Matrix (JDM) is given by the number of edges between nodes of degree $i$ ($V_i$) and $j$ ($V_j$) as in \cite{amanatidis2015graphic}:
\begin{equation}
JDM(i,j) = \sum_{u \in V_i}\sum_{v \in V_j}1_{\{(u,v)\in E\}}
\end{equation}

Realizability conditions were given by \cite{amanatidis2015graphic} and several other algorithms to produce single realizations for given JDM \cite{czabarka2015realizations}, \cite{gjoka2015construction} and \cite{stanton2012constructing}. The algorithms presented in \cite{amanatidis2015graphic} and \cite{stanton2012constructing} are designed to only produce more restricted realizations with a property called Balanced Degree Invariant, while in \cite{czabarka2015realizations} and \cite{gjoka2015construction}, the algorithms have non-zero probability to produce any realization of a 2K distribution. An importance sampling algorithm was introduced by Bassler et. al. \cite{bassler2015exact}. 

Similarly an extension of JDM, Joint Degree and Attribute Matrix (JDAM) is given in graph with a single node attribute by the number of edges between nodes of degree $i$, attribute value $p$ ($V_{\{i,p\}}$) and degree $j$, attribute value $q$ ($V_{\{i,q\}}$) as in \cite{gjoka2015construction}. Known results can be easily applied from the JDM problem including sampling.
\begin{equation}
JDAM(\{i,p\},\{j,q\}) = \sum_{u \in V_{\{i,p\}}}\sum_{v \in V_{\{j,q\}}}1_{\{(u,v)\in E\}}
\end{equation}
{\bf dK, $d>2$ Construction.} While several attempts were made to find polynomial time algorithms to produce 3K graphs \cite{mahadevan2006systematic} or 2K realization with prescribed (degree-dependent) clustering coefficient in \cite{gjoka20132},\cite{gjoka2015construction} and \cite{orsini2015quantifying}, it was recently proved that even the realizability check for these inputs is NP-Complete in \cite{devanny2016computational}.

Annotated graph construction was proposed in \cite{Dimitropoulos:2009:GAM:1596519.1596522} that considered degree correlations, however the proposed construction method will generate graphs with self-loops or multi-edges initially. An additional step removes these extra edges to make the graph simple and finally the largest connected component of the graph is returned as the constructed realization.

The space of simple realizations of 1K distributions is connected over double edge swaps preserving degrees \cite{taylor1980constrained} and a similar result was shown in \cite{czabarka2015realizations} or \cite{amanatidis2015graphic} for 2K with double edge swaps that preserve degrees and the joint degree matrix. These swaps allow the use of MCMC to generate approximate probability samples for 1K and 2K. However, fast mixing for the MCMC has not been proved in general, only for special classes of realizations \cite{erdHos2016new} \cite{erdos2015decomposition}.

We highlight related work for bipartite and directed degree sequences in details in the following section as part of our description of directed dK-series.

\begin{figure*}[t!]
\includegraphics[scale=0.34]{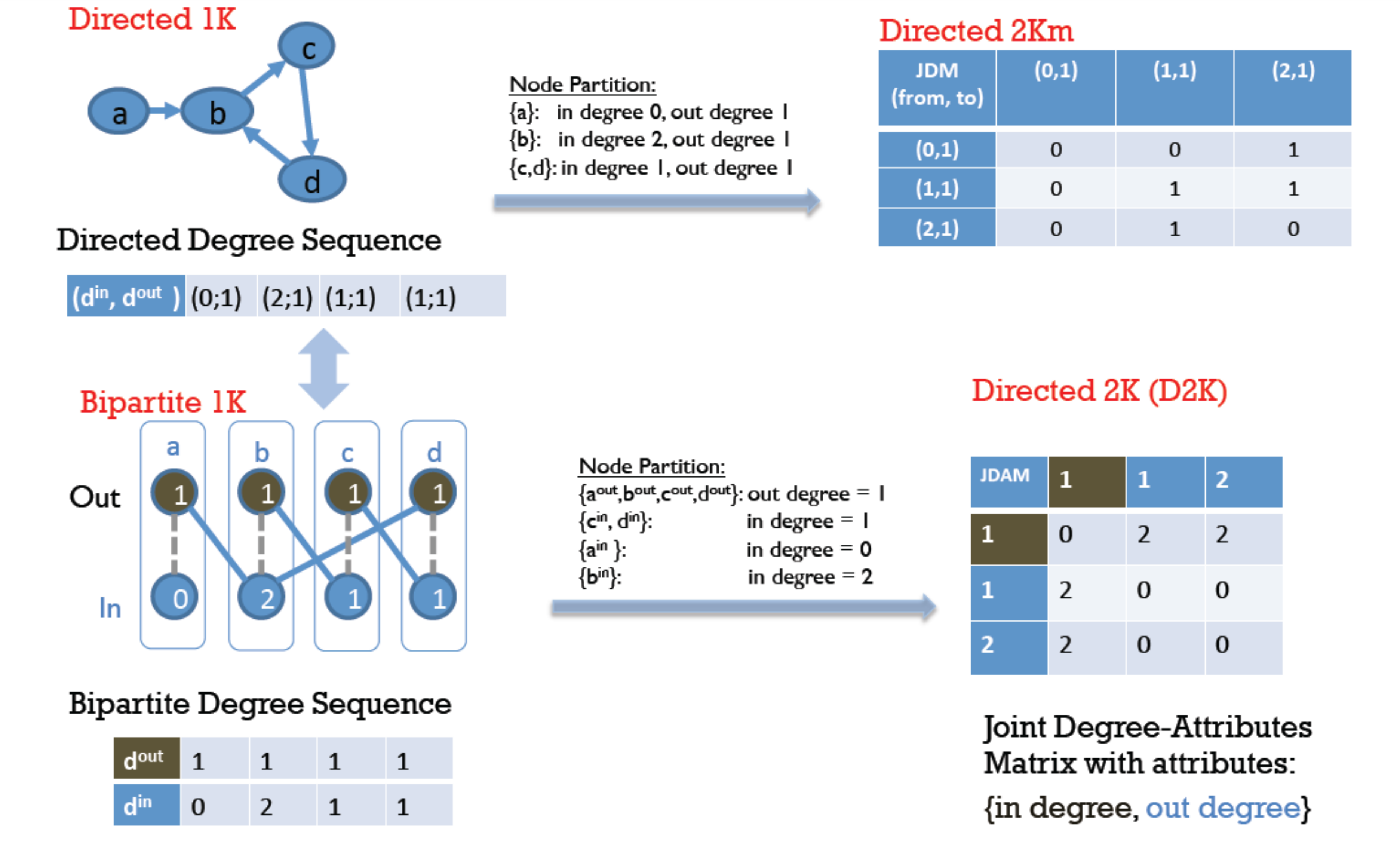}
\caption{Defining Directed 2K, to capture degree correlations in a directed graph. {\bf Top left, Directed 1K:} Directed graph with a given degree sequence (DDS). {\bf Bottom left, Bipartite 1K:} Mapping of the previous to a Bipartite undirected graph with a given  bipartite degree sequence; non-chords in the bipartite graph (shown in dashed line) correspond to self-loops in the directed graph.   {\bf Bottom right, Directed 2K (D2K):} Joint-Degree-Attribute Matrix (JDAM), where nodes of the bipartite graph are partitioned by their degree-and-(in or out) attribute.
{\bf Top right: Directed 2Km:} Joint Degree Matrix (JDM) for directed graphs, where nodes are partitioned according to their (in degree, out degree).}\label{fig:problems} 
\end{figure*}

\section{Directed 2K Problem Definition}\label{sec:problem}

We are able to define analogous distributions for the dK-series to capture directed graphs with given properties. 

{\bf Directed 0K.} As mentioned in Section \ref{sec:related}, it is trivial to extend ER model for directed graphs. In addition, we consider another model called UMAN \cite{holland1976local}, that captures the number of mutual, asymmetric, and null dyads in a graph. We can think about UMAN as 0K distribution with fixed numbers of mutual and unreciprocated edges.

{\bf Directed 1K.} In an undirected graph $G$, a node $v$ has degree $d(v)$, and the degree sequence is simply \break $DS=\{d_1, d_2, ... d_{|V|}\}$. In a directed graph, a node $v$ has both in and out degree and the {\em directed degree sequence} is $DDS=\{(d^{in}_v, d^{out}_v), v\in V\}$; an example is shown on Fig. \ref{fig:problems}-top left.

It is well known from Gale's work \cite{gale1957theorem}, that any directed graph can be 1-1 mapped to an undirected bipartite graph, where each node $v$ of the directed graph is split in two nodes $v_{in}$ and $v_{out}$, and the undirected edges across the two (in and out) partitions of the bipartite graph correspond to the directed edges in the directed graph. A self loop $(v,v)$ in the directed graph  corresponds to a ``non-chord''$(v_{in}, v_{out})$  in the bipartite graph, and is shown in dashed line on  Fig. \ref{fig:problems}-left-bottom. 

Construction algorithms are known for a bipartite degree sequence with \cite{gale1957theorem},  or without non-chords, and therefore of the corresponding directed graphs with or without  \cite{fulkerson1960zero} self loops, respectively. More recently, an importance sampling algorithm was shown in \cite{kim2012constructing}.

{\bf Directed 2K.}
 Our goal in this paper is to go beyond directed degree sequence and capture degree correlation, and there are two ways to go about it.

\noindent {\bf D2Km: Joint (Directed) Degree Matrix \break $JDM((d^{in}_i, d^{out}_i),(d^{in}_j, d^{out}_j))$}.

One way is to work directly with the directed graph, see Fig. \ref{fig:problems}-top row. We can partition nodes by both their in and out degrees $(d^{in}_v, d^{out}_v)$, and we can define the joint degree matrix to capture the number of edges $JDM((d^{in}_i, d^{out}_i),(d^{in}_j, d^{out}_j))$, between nodes with  $(d^{in}_i, d^{out}_i)$ and  $(d^{in}_j, d^{out}_j)$. This is shown on the top of Fig. \ref{fig:problems}. This is a natural extension of the JDM in the undirected case, and expresses a restrictive notion of degree correlation.  However, there is no known algorithm that can provably generate this target JDM. For example our own 2K construction algorithm for undirected graphs \cite{gjoka2015construction} does not work all the time, although it is still a good heuristic for sparse graphs.

\noindent {\bf  D2K: Joint Degree-Attribute Matrix $JDAM(degree,$ $in$ $or$ $out)$}

An alternative approach is to work with the equivalent representation as an undirected bipartite graph without non-chords (Fig. \ref{fig:problems}-bottom left), and define degree correlation there. We partition in and out nodes by their degree, essentially considering that nodes in the bipartite graph can have an attribute that takes two values, ``in'' or ``out.'' We can now define degree correlation using the Joint Degree-Attribute Matrix (JDAM, which we first defined in \cite{gjoka2015construction}), as shown on Fig. \ref{fig:problems}- bottom right.  This leads to a  JDAM with two attribute values, such that $\forall i,j = 1,...,d_{max}$ degrees and $p \in \{in,out\}$ attribute values $JDAM(\{i,p\},\{j,p\})=0$, {\em i.e.,} because the bipartite graph has no edges between two ``in'' or ''`out'' nodes. Furthermore, the number of non-chords will be noted as $f(\{i,p\},\{j,q\})$, where $i,j\ \in \{1,...,d_{max}\}$ and $p \neq q \in \{in,out\}$; $f$ can be computed by passing through the directed degree sequence once and counting the number entries with in-degree $i$ and out-degree $j$.

This notion of $bipartite$ $JDAM$ has all the properties known for $JDM$ and $JDAM$, since it is a special case of $JDAM$ that we first defined in \cite{gjoka2015construction}. This includes sufficient and necessary conditions for realizability, construction algorithms, existence of Balanced Degree Invariant realizations, importance sampling algorithm extensions from $JDM$, connectivity of space of realizations over $JDAM$ preserving double-edge swaps and MCMC properties. However, we have to show for D2K, that the non-chords described by the directed degree sequence can be added as well.

{\bf Relation of the two problems.} An overview of the problems of interest is provided on Fig. \ref{fig:problems}. One difference between the two 2K problems  is that D2Km provides a more restrictive notion of degree correlation than D2K since it partitions nodes by two numbers $(d^{in}, d^{out})$) vs. one number  $d^{in}_v$ or $d^{out}_v$. D2Km can essentially be obtained as a special case of D2K by further partitioning nodes with the same  $d^{in}$  by their out degree as well. Therefore, D2Km can be solved by the same algorithm that solves D2K. For the rest of the paper, we will consider the D2K problem.

\medskip
\noindent {\bf Directed 2K (D2K) Problems:} Given targets  JDAM {\em and} DDS:
\begin{itemize} 
\item {\bf Realizability:} Decide whether this D2K is realizable, {\em i.e.,} whether there exist graphs with these properties.
\item {\bf Construction:} Design an algorithm that constructs at least one such realization.
\item {\bf Sampling:} Sample from the space of all graph realizations with the target D2K.
\end{itemize}

\section{Realizability and Algorithm}  \label{sec:algorithm}

In this section, we take as input the two target properties, namely the target $\target{\JDAM(\{i,p\},\{j,q\})}$ with two attribute values and the $\target{\DDS}$, and construct a directed 2K-graph with $N$ nodes that exhibits exactly these  target properties. In this section, we use the bipartite representation of directed graphs as in Figure \ref{fig:problems}; this enables us to simplify our algorithm description, loose directionality of the edges and only handle a partition with non-chords. 

Recall that in the D2K definition, nodes are partitioned into $K$ parts $V_k, k=1...K$, according to the distinct $d^{in}=i$ or $d^{out}=j$ they exhibit and $\JDAM(\{i,p\},\{j,q\})$ is indexed accordingly. For example,  on Fig. \ref{fig:problems} bottom-right, each node belongs to one of four parts $V_{\{0,in\}}=\{v\in V: d^{in}=0\}$, $V_{\{1,out\}}=\{v\in V: d^{out}=1\}$, $V_{\{1,in\}}=\{v \in V: d^{in}=1\}$, $V_{\{2,in\}}=\{v \in V: d^{in}=2 \}$, and the JDAM is 3x3 (by removing rows and columns corresponding to any $V_{\{0,p\}}$, since there are no edges using these parts of any partition).

\subsection{Realizability}

Not all target properties are realizable (or ``graphical''): there does not always exist at least one simple directed graph with those exact properties. Necessary and sufficient conditions for a target D2K, {\em i.e.,}  $\target{\JDAM(\{i,p\},\{j,q\})}$ and $\target{\DDS}$, to be realizable are the following.

\begin{itemize}
\item[I] $\forall i,j,p: JDAM(\{i,p\},\{j,p\})=0$
\item[II] $\forall i,j,p,q$, if $JDAM(\{i,p\},\{j,q\})>0$, \break $JDAM(\{i,p\},\{j,q\})$ $+ f(\{i,p\},\{j,q\})\leq |V_k| \cdot |V_l|$
\item[III] $\forall i,p: |V_{\{i,p\}}| = \sum_{\{j,q\}} \frac{JDAM(\{i,p\},\{j,q\})}{i}=$ number of times $i$ appears in $DDS$ as $p$ and it is an integer.
\end{itemize}

These are generalizations of the conditions for an undirected JDM, JDAM to be realizable, and they are clearly necessary. The first condition states that the target JDAM is bipartite, \ie~there should be no edges between two nodes both in ``in'' or ``out'' parts. The second condition considers edges between two (``in'' and ``out'' ) parts and states that the number of edges defined by the $JDAM(\{i,p\},\{j,p\})$ plus the number of non-chords should not exceed the total number of edges possible in a complete bipartite graph across the two parts. The last condition ensures that the target \JDAM and the target \DDS{} are consistent: the number of nodes with in (or out) degree $i$ should be the same whether computed using the \JDAM or the \DDS. The conditions are shown to be sufficient via the constructive proof of the algorithm.
Necessity of these conditions for simple graph construction are trivial.

\begin{figure*}
\centering
\includegraphics[height=1.7in, width=7in]{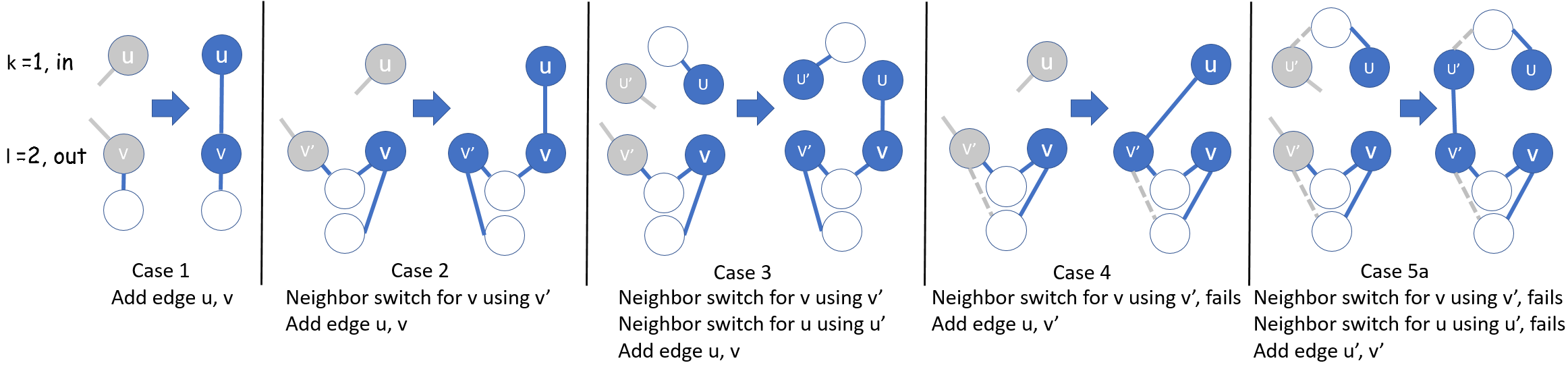}
\caption{Different possible cases, while attempting to add $(u,v)$ edge in Algorithm 1.}
\label{fig:proofcases}
\end{figure*}

\subsection{Algorithm}

\begin{minipage}{\linewidth}
\vspace{5pt}
\hangindent=9pt
Algorithm 1\\
Input: $\target{DDS}$, $\target{JDAM}$ \\
Initialization: \\
a: Create nodes, partition, stubs using $\target{DDS}$ \\
b: Add non-chords to G using $\target{DDS}$\\
Add Edges:\\
1: \FOR every pair $(\{i,p\},\{j,q\})$ of partition: \\
2: \SP \WHILE $JDAM(\{i,p\},\{j,q\})<\target{JDAM}(\{i,p\},\{j,q\})$ \\
3: \SP \SP Pick any nodes $u$ (from $V\{i,p\}$), $v$ (from $V\{j,q\}$) \\
   \SP \SP \SP s.t. $(u,v)$ is not a non-chord or existing edge\\
4: \SP \SP \IF $u$ does not have free stubs: \\
5: \SP \SP \SP $u'$: node in $V\{i,p\}$ with free stubs \\
6: \SP \SP \SP neighbor switch for $u$ using $u'$ \\
   \SP \SP \SP \SP \SP \IF neighbor switch fails, $u := u'$ \\
7: \SP \SP \IF $v$ does not have free stubs: \\
8: \SP \SP \SP $v'$: node in $V\{j,q\}$ with free stubs \\
9: \SP \SP \SP neighbor switch for $v$ using $v'$ \\
   \SP \SP \SP \SP \SP \IF neighbor switch fails, $v := v'$ \\   
10: \SP \SP add edge between ($u$, $v$) \\
11: \SP \SP $JDAM(\{i,p\},\{j,q\})$ ++;  $JDAM(\{j,q\},\{i,p\})$ ++; \\
12: Transform bipartite G to directed graph \\
Output: simple directed graph G\\
\end{minipage}

First, we create a set of nodes $V$, where $|V|=2\cdot|\DDS|$, we assign stubs to each node and partition nodes, as specified in the \emph{target directed degree sequence} $\target{\DDS}$. The stubs are originally free, {\em i.e.,} they are only connected to one node. We also initialize all entries  of \JDAM to 0 and the non-chords between nodes according to $\target{\DDS}$.
Then the algorithm proceeds by connecting two nodes (one from ``in'' and one from ``out'' side), thus adding one edge $(u,v)$ at a time, that (i) are not previously connected to each other (ii) do not have a non-chord between them (to avoid self loops) and (iii) for whom the corresponding entry in the \JDAM  has not reached its target. The challenge lies in showing that the algorithm will always be able to make progress, by adding one edge at a time, until  all entries of the \JDAM reach their target, when the algorithm terminates. Indeed, there may be cases (2-5  in Fig.\ref{fig:proofcases}), where $u$, $v$ or both do not have free stubs. Even in those cases, however, we will be able to perform JDAM-preserving edge rewirings (called neighbor switch  \cite{gjoka2015construction}: remove a neighbor $t$ of $v$ such that $t$ is not a neighbor of $v'$ and add edge $(t,v')$) to free stubs and then add the edge $(u,v)$ (cases 2-3); or we will be able to add another edge $(u',v')$ (cases 4-5a). Next, we prove that this is indeed always the case.

\begin{proof}$\{$Here, we follow the style of proof from Gjoka et. al. \cite{gjoka2015construction}. However, we would like to point out that other similar results for JDM construction could be extended for directed 2K as well, such as the proof in \cite{czabarka2015realizations}.$\}$ 

Condition II. guarantees that two nodes can be always chosen to add an edge and Condition III. ensures that at least one node exist in every part of the partition as long as $JDAM(i,j) < \target{JDAM}(i,j)$\footnote{For more details please read Lemma 2. and 3. in \cite{gjoka2015construction}.}. Now, we simply show that every iteration can proceed by adding a new edge to the graph:

{\em Case 1.} Add a new edge between two nodes w/free stubs, no local rewiring needed. 

{\em Case 2.}  Add a new edge between a node $v$ w/out free stubs and a node $u$ w/free stubs where neighbor switch is possible for $v$ without using any non-chords.

{\em Case 3.} Add a new edge between two nodes w/out free stubs where neighbor switches are possible for both nodes without using any non-chords.

{\em Case 4.} Add a new edge between a node $v$ w/out free stubs and a node $u$ w/free stubs (or w/out free stubs where neighbor switch is possible) where neighbor switch is not possible for $v$ using $v'$ without using any non-chords. In this case $v'$ has the same neighbors as $v$ except the one for which it has an assigned non-chord. In this case $v'$ is not connected to $u$ and it is possible to add $\{v',u\}$ edge ($\{v',u\}$ is clearly not an edge since then $u$ would be also connected to $v$ or $v$ could have done a neighbor switch).

{\em Case 5.} Add a new edge between two nodes ($u,v$) w/out free stubs, where neither can do a neighbor switch with $u'$ and $v'$ respectively. We have to break this case into two subcases, based on whether two nodes $u',v'$ w/out free stubs form a non-chord or not.

{\em Case 5a.} $u',v'$ is not a non-chord. This means that we can add a new edge between $u',v'$. It is easy to see that $u',v'$ edge is not already present, because otherwise $u$ and $v$ could have performed a neighbor switch.

{\em Case 5b.} $u',v'$ is a non-chord. This case is not possible when $u,v$ are not able to perform neighbor switches at the same time. Without loss of generality, let's say that $u$ connects to all the neighbors of $u'$ and node $v'$. This means that no neighbor switch is available for $u$. Now, if we want to construct $v$ such that it can't perform a neighbor switch with $v'$, we need $v$ to connect all of the neighbors of $v'$; however, this would include $u$ as well, and clearly that edge doesn't exist. Contradiction.

This concludes our proof and shows that the algorithm will terminate and generate a bipartite graph after adding $\sum{\target{JDAM}(i,j)}/2=|E|$ edges.
\end{proof}

{\bf Running Time.} The time complexity of the above algorithm is $O(|E|\cdot d_{max})$. In each iteration of the {\tt while} loop, one edge  is always added, until we add all $|E|$ edges.  However, we have to consider how much time it takes to pick such nodes. There could be neighbor switches that remove previously added edges or add edges between the two parts. If we naively looked up node pairs, it would become an issue for dense graphs. A simple solution is to keep track of $\target{JDAM}(\{i,p\},\{j,q\})- JDAM(\{i,p\},\{j,q\})$ many node pairs where edges can be added in a set $P$. For every pair of  $\{i,p\},\{j,q\}$, it is possible to initialize $P$ by passing through $O(\target{JDAM}(\{i,p\},\{j,q\})+ f(\{i,p\},\{j,q\}))$ node pairs. A new $(u,v)$ node pair can simply be picked as  a random element from $P$. If a neighbor switch for $u \in V_{\{i,p\}}$ (and similarly to $v$), rewires a neighbor $t \in V_{\{j,q\}}$, then $P= P \setminus \{u',t\}\cup \{u,t\}$  maintains available node pairs in $P$. Note: $\{u',t\}$ might not be in $P$. This ensures that $|P|\geq JDAM(\{i,p\},\{j,q\})- JDAM'(\{i,p\},\{j,q\})$. These simple set operations can be done in constant time, and building $P$ takes $O(E+V)$ time over all partition class pairs. Finally we remove $(u,v)$ from $P$, which could be different from the starting pair if Case 4 or 5a occurs.

It is also possible to keep track of nodes with free stubs in a queue for each part of the partition. Once a node has no free stubs, it will remain so, except {\em during} neighbor switches. This allows selection of candidates for neighbor switches, or new edges when neighbor switches are not possible, in constant time. However, it still takes $O(d_{max})$ to check whether a node with free stubs is a good candidate for neighbor switch, because the sets of neighbors can be almost the same length, which takes linear time in the size of sets. In the worst case, there is a possibility for at most two neighbor switches per new edge, hence the running time is $O(|E| \cdot d_{max})$.

The directed graph can be constructed from the bipartite representation by collapsing nodes with non-chords and assigning directions to edges appropriately, this takes $O(E+V)$ time.

\subsection{Space of Realizations}
\label{sec:space}

The order in which the algorithm adds edges is unspecified. The algorithm can produce any realization of a realizable D2K, with a non-zero probability. Considering all  possible edge permutations as the order in which to add the edges, the ones where no neighbor switch is required correspond to all the possible realizations. Unfortunately, the remaining orderings are difficult to quantify, thus the current algorithm cannot sample uniformly from all realizations with a target D2K during construction.

Another  way to sample from the space of graph realizations with a given D2K is by edge rewiring. This method is typically used by MCMC approaches that transform one realization to another by rewiring edges so as to present the target properties. On the positive side, D2K is a special case of an undirected JDAM, and thus inherits the property that JDAM realizations are  connected via 2K-preserving double-edge swaps \cite{czabarka2015realizations},\cite{amanatidis2015graphic} if non-chords are allowed (equivalently, self-loops in directed graphs). On the negative side, we cannot use the known swaps to sample from the space of simple directed graphs. 

\begin{figure}
\centering
\includegraphics[height=1.1in, width=2in]{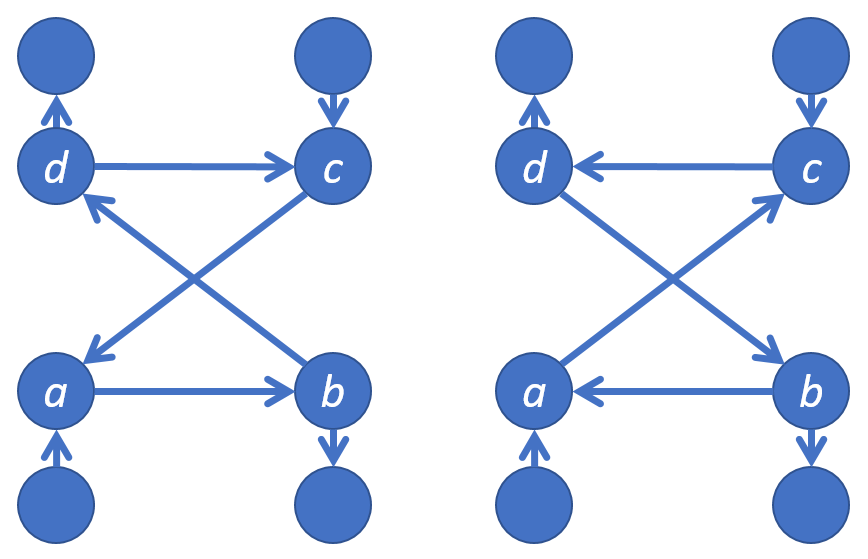}
\caption{Two realizations with the same degree sequence and JDAM. There is no JDAM preserving double-edge swap that would not use any self-loops and the $C_6$ swaps are not preserving JDAMs. This shows that the edges along the 4-directed-cycle must change their direction simultaneously.}
\label{fig:complexswap}
\end{figure}

The space of simple realizations of directed degree sequences is connected over double edge swaps, that preserve (in and out) degrees, and triangular $C_6$ swaps. Triangular $C_6$ swaps are necessary in some cases where the difference between two realization is the orientation of a directed three-cycle: in this case, the orientation of the cycle has to be reversed in a single step. The sufficiency of only these two types of swap was shown in \cite{erdHos2013swap}. The necessity of these swaps also carries over to (simple) directed 2K realizations. However, Fig. \ref{fig:complexswap} shows a counterexample (a directed 4-cycle) where the classic swaps are not sufficient to transform one realization to the other, thus requiring a more complex swap. We leave it as an open question whether tight upper bounds can be derived on the swap size for Directed 2K realizations.  \footnote{There are possibly other cases where swaps must be more complex and include more edges at once, for example larger directed cycles with specific in/out degree order. In this paper, we do not provide tight upper bounds on the number of self-loops or the size of swaps required, but we do emphasize that no multi-edges are required and the number of self-loops are of course bounded by $|N|$.}

\section{Evaluation on Real-World Graphs} \label{sec:results}

\subsection{Datasets}
We have used examples of directed graphs for our experiments from SNAP \cite{snapnets}: p2p-Gnutella08, Wiki-Vote, AS-Caida, Twitter. We have chosen these networks in order to represent several different sizes and generating processes for directed graphs.

We have removed any present self-loops or multi-edges from these graphs, since our goal is to produce simple graphs and this step ensures that the measured inputs from these graphs will be realizable using different directed graph construction methods (0K, UMAN, 1K, 2K, 2Km). 

AS-Caida has edge labels according to the relationship between two ASes (peer, sibling, provider, customer), however provider and customer edges describe the same relation from the opposite point of view. We have modified the AS-Caida network by removing customer relations between ASes. The affect of the mutual edges - which would be present using both provider and customer relations - will be visited again in our discussion. In \cite{Dimitropoulos:2009:GAM:1596519.1596522}, this graph was also considered with the relations included, however during our construction, we only use directed dK-series as described before in Section \ref{sec:problem}.

An overview of the final graphs used in our experiments is shown in Table \ref{tab:graphs}.

\begin{table}
\centering
\caption{Graphs}
\begin{tabular}{|c|c|c|} \hline
Name&\#Nodes&\#Edges\\ \hline
p2p-Gnutella08 & 6,301 &  20,777 \\ \hline
Wiki-Vote & 7,115 &  103,689 \\ \hline
AS-Caida & 26,475 &  57,582 \\ \hline
Twitter & 81,306 &  1,768,135 \\ 
\hline\end{tabular}
\label{tab:graphs}
\end{table}

\subsection{Properties}
In our results, we are going to evaluate the performance of graph generators in terms of properties associated with directed graphs. While the size of the generated graphs are maintained (number of nodes and edges), there are many other properties one could investigate.

To evaluate correctness of our implementation of D2K, first we use Degree Distributions and Degree Correlations. These are the ones expected to be exactly matched by definition.

In addition, we consider additional properties such as shortest paths, spectrum, strongly connected components, betweenness centrality, and k-core distribution. Finally, to measure how some of the local structures are preserved, we use dyad and triad censuses, dyad-wise shared partners, average neighbor degree, and expansion.

The dyad census counts the different configurations for every pair of nodes: ''mutual" - edges in both direction, ''asymmetric" - edge only in one direction and ''null" - no edge present. The triad census counts the non-isomorphic configuration for every triplet of nodes. A complete list of configurations and naming conventions can be found in \cite{holland1976local}. Configurations are identified by three numbers (mutual, asymmetric, and null counts) and a letter in case of different non-isomorphic configuration with the same number of edges. For example ''003" is a triplet of nodes where none of the edges are present, ''030C" is a directed 3-cycle and ''300" is a triplet of nodes where all directed edges are present.

Shared partners for pairs of nodes can be defined in three ways for directed graphs: using independent two-paths, using shared outgoing neighbors or using shared incoming neighbors between pairs of nodes \cite{snijders.et.al:sm:2006}. Dyad-wise shared partners (DSP) count node pairs by the number of shared partners appearing in a network.

Average neighbor degree captures the average degree of a nodes' neighbors, and we split this property for in - and out degrees. Similarly, we refer to \emph{expansion} for directed graphs as the ratio of the first hop and second hop neighborhoods' sizes going out, or coming in to a node. These properties capture similar aspects of a network, but expansion excludes any mutual edges or edges between nodes in the first hop neighbors.

Some of above properties are also used by Orsini et. al. \cite{orsini2015quantifying} to study the convergence of dK-series over different types of undirected networks. However, we have included a few properties more natural to discuss for directed graphs, such as the triad census.

\subsection{Implementation}

While our current code is not available online, the undirected version of our algorithm is available in NetworkX \cite{hagberg-2008-exploring} and easy to modify for D2K or D2Km using the description in Section \ref{sec:algorithm}. We plan to release our code for D2K and D2Km as part of the NetworkX library in the near future. Until then, please contact the first author for the implementation. 

The implementation of the used graph properties is available as part of NetworkX or trivial to implement using the above description.

\subsection{Results}
We compare realizations generated by Directed ER (D0K), UMAN, Directed Degree Sequence (D1K), Directed 2K, Directed 2Km with the corresponding target properties captured on input graph (G). We use 20 random instances for every construction method and then average our results for each specific property.

\begin{figure*}
\centering
\includegraphics[height=2.02in, width=7in]{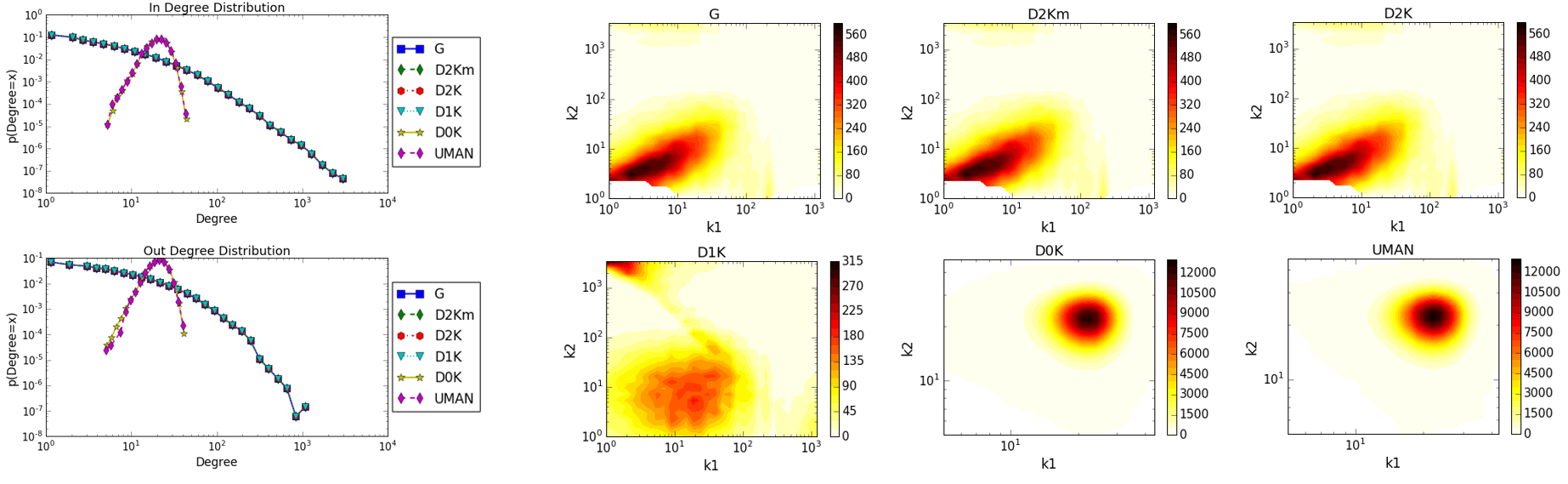}
\caption{Results for Twitter graph: Directed Degree Distribution and Degree Correlation}
\label{fig:results_a}
\end{figure*}

\begin{figure*}
\centering
\includegraphics[height=2.02in, width=7in]{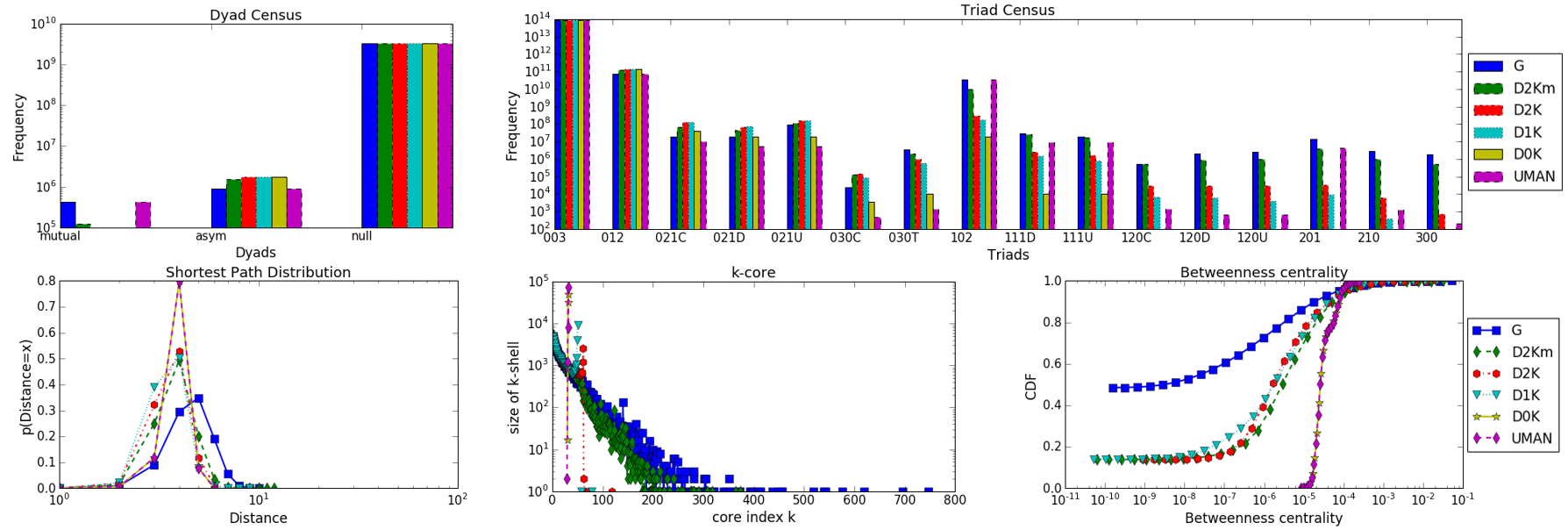}
\caption{Results for Twitter graph: Dyad-, Triad Census, Shortest Path Distribution, K-core distribution, Betweenness Centrality}
\label{fig:results_c}
\end{figure*}

\begin{figure*}
\centering
\includegraphics[height=2.02in, width=7in]{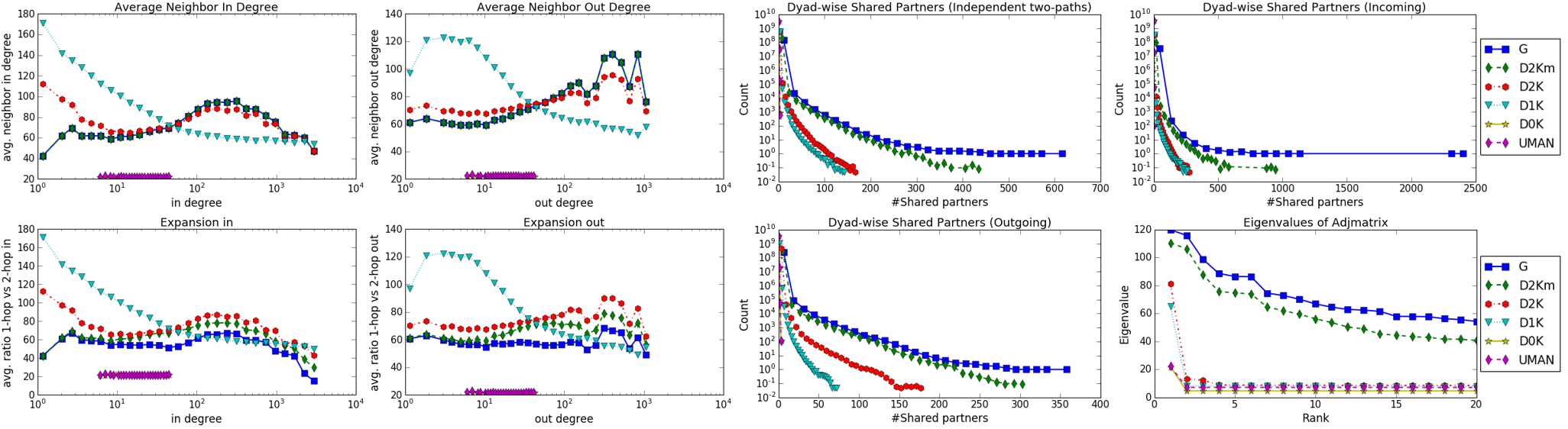}
\caption{Results for Twitter graph: Expansion, Average Neighbor Degree, DSP and top 20 Eigenvalues}
\label{fig:results_b}
\end{figure*}

Due to space constraints, we provide detailed results only for the Twitter graph in Figure \ref{fig:results_a}, \ref{fig:results_c} and \ref{fig:results_b} and a brief overview of our observations over different input graphs. 

First, we can observe in Figure \ref{fig:results_a} that Directed Degree Distributions and Degree Correlations are captured by D2K, D2Km as expected by definition. This shows that our implementation is correct. On the other hand, D0K, D1K and UMAN  capture Degree Correlations poorly, thus D2K graphs have a possibility to capture other properties more accurately than D0K or D1K.

{\bf Dyad Census} is not well captured for Twitter, as we can see in Figure \ref{fig:results_c}. However, there are order of magnitude improvements in the number of mutual edges between D2Km (123,040.4) and D2K (3,628.7), D1K (2,155.95) or D0K (233.05). Of course, UMAN preserves this property by definition. 

{\bf Triad Census} is surprisingly well captured by UMAN, the reason being the exact match for the Dyad Census in the previous point. On the other hand, a convergence can be seen between dK-series generators with significant improvements in dense triadic structures from D1K to D2K and from D2K to D2Km.

{\bf Betweenness Centrality CDF} has no significant improvements after matching degree distributions with D1K in Twitter; other examples reached target closer with D1K. Interestingly UMAN performs almost identically to D0K, even though the number of mutual edges is significantly different.

{\bf Shortest Path Distribution} has slow convergence to target across different methods, but the average shortest path is shorter than the observed in $G$.

{\bf Strongly Connected Components (SCC)} are not well captured by any of the dK-series generators and they tend to produce realizations with a single giant SCC and many one-node components without any intermediate sizes of SCCs.

{\bf K-Core Distribution} is best captured by D2Km, and there is a small improvement from D1K to D2K using Twitter. However, the dense core using D1K or D2K is almost an order of magnitude lower core index.

{\bf Eigenvalues} of Twitter is again best targeted by D2Km. There is a difference between leading eigenvalues in graph realizations of the other methods but starting at the second eigenvalue the difference between D1K and D2K quickly decreases.

{\bf Dyad-wise Shared Partners} follow similar trends to other properties, such that D2Km is significantly more accurate than D1K and D2K. D2K improves over D1K in terms of ''outgoing shared partners" but that improvement decreases at ''independent two-paths" and disappears at ''incoming shared partners".

{\bf Expansion} property is again best approximated by D2Km and D2Km even matches {\bf Average Neighbor Degree} exactly if marginalized by degrees as in Figure \ref{fig:results_b}. D2K also follows the general shape of these distributions but includes larger error, while D1K has systematic difference compared to $G$.

\begin{table}
\centering
\caption{Summary of results: showing improvements by fixing more properties. Labels: ''." - no improvement, ''-" - decreased accuracy, ''+" - increased accuracy, ''Exact" - matched by definition.}
\begin{small}
\begin{tabular}{|c|c|c|c|} \hline
Property & UMAN$\rightarrow$D1K  & D1K$\rightarrow$D2K &  D2K$\rightarrow$D2Km\\ \hline
Degree Distribution  & Exact & Exact & Exact \\ \hline
Degree Correlation  & + & Exact  & Exact \\ \hline
Dyad Census  & -& +& + \\ \hline
Triad Census & +& +&+ \\ \hline
Betweenness Centrality & +& . & . \\ \hline
Shortest Path Distribution & +& +& + \\ \hline
Eigenvalues  & +& + & + \\ \hline
DSP  &+&+&+ \\ \hline
Expansion & + & +& + \\ \hline
Avg. Neighbor degrees & +& +& Exact \\ \hline
S. Connected Components & .&. &. \\ \hline
K-Core Distribution  & + & . & + \\
\hline\end{tabular}
\end{small}
\label{tab:results}
\end{table}

Table \ref{tab:results} gives an overview of how network properties are affected by the different dK graph construction methods for the other remaining networks. The Twitter network showcased most of our general findings, but individually some of these networks have characteristics that makes them different from Twitter, e.g. p2p-Gnutella08 does not contain any mutual edges. The most interesting question is whether D2K or D2Km capture network properties more accurately. The answer is yes in most cases, but it might not be a significant improvement in targeting certain properties.

Local structures are generally better captured by D2K and even more precisely for D2Km, but global properties might not be significantly affected depending on the original network. However, this result is not surprising, since one of the main assumptions of the dK-series is that it is not necessary to target high $d$ values for every graph \cite{orsini2015quantifying}.

\subsection{Discussion}
We have used several construction algorithms to generate random graphs and we have observed convergence for most properties in the directed dK-series. However, certain properties are not well captured by these methods, such as the Strongly Connected Components. This shows that further extensions or heuristics would be practical to extend the directed dK-series. 

For some networks D1K could be a good choice, since it captures many network properties. However, we have shown that even in those cases properties related to first hop neighbors are not expected to be captured.

While for most metrics the degree labeled construction performs reasonably well -starting from D1K- it is important to notice that these methods create a low number of mutual edges. On one side, UMAN generates graphs with prescribed number of mutual edges and otherwise ER random graph-like structures. On the other hand, for larger (sparse) networks the degree labeled construction only achieves a fraction of the target number of mutual edges. A solution to this problem would be to generate D2K graphs with a number of mutual edges. It is possible to design heuristics for this problem but exact solutions might be difficult to achieve. In addition, we could consider for D2Km two matrices: one describing asymmetric and another for mutual edges between nodes with given (in, out) degrees.

We have also shown that further partitioning of nodes can help to better describe graphs by their mixing. D2Km is a very specific partition, that preserves average neighbor degrees for directed graphs, which is a property that is given by 2K in the undirected case. In the limit of the possible partitions the graphs could be exactly fixed; however, in our example this is not the case. While there are parts of the partition with only a single node in them, most of their edges go to other parts of the partition with multiple nodes. This results in a chance to construct distinct realizations with minimum number of fixed edges across different realizations. It is an interesting question, how different partitions would affect the number of realizations. We leave it as future work for both undirected and directed cases.

The limitations of degree labeled construction for undirected graphs have been shown recently, such that the realizability problem of undirected 2K with fixed number of triangles, 3K \cite{devanny2016computational}, and second order degree sequences (degree and number of two hop neighbors) \cite{erdHos2016not} are NP-Complete and a relaxation of JDM partitions called PAM \cite{erdHos2015graph} is believed to be NP-Complete and only solved for special cases. We can say with good confidence that more restrictive models are likely to lead to NP-Complete problems; however, different partitions can be solved as long as they produce a valid JDAM for the D2K problem.

\section{Conclusion}
\label{sec:conclusion}
We have shown a new approach for directed graph construction by considering in and out degree correlations in addition to directed degree sequences. This extension enabled us to build a framework for directed graphs similar to the dK-series, using bipartite graphs. We solve the problem of generating bipartite graphs with prescribed degree correlations using the Joint Degree and Attribute Matrix construction algorithms. Following in the footsteps of classic work, we use this property of the JDAM problem and defined directed 2K as the combination of JDAM and a directed degree sequence. To solve our proposed problem, we provide the necessary and sufficient conditions for realizability of such inputs and a simple, efficient algorithm to generate simple realizations (without self-loops) as well. The uniform sampling from the space of these realizations is not trivial, as we have shown an example for a necessary complex edge swap in section \ref{sec:space}. However, we see an opportunity for further research in edge swaps for an MCMC approach or a variant of the importance sampling algorithm from \cite{bassler2015exact}.

In addition to directed 2K, we have defined D2Km, a special case using additional node attributes. D2Km provides a more restricted notion of directed 2K, that exactly captures average neighbor degree of nodes marginalized by degree.  In our experiments, we have shown convergence for degree labeled directed graph construction similar to the undirected case \cite{orsini2015quantifying}. This result is similar to \cite{orsini2015quantifying} in nature but it shows that degree correlations capture more information about graph structure and enables us to generate graphs which more closely resemble real-life networks.

We have identified that directed dK-series can benefit directly by prescribing the number of mutual edges. We consider this extension the most straightforward step to improve directed dK-series. However, even this problem could turn out to be NP-Complete.

\bibliographystyle{ACM-Reference-Format}
\bibliography{references}  


\begin{thebibliography}{00}


\ifx \showCODEN    \undefined \def \showCODEN     #1{\unskip}     \fi
\ifx \showDOI      \undefined \def \showDOI       #1{{\tt DOI:}\penalty0{#1}\ }
  \fi
\ifx \showISBNx    \undefined \def \showISBNx     #1{\unskip}     \fi
\ifx \showISBNxiii \undefined \def \showISBNxiii  #1{\unskip}     \fi
\ifx \showISSN     \undefined \def \showISSN      #1{\unskip}     \fi
\ifx \showLCCN     \undefined \def \showLCCN      #1{\unskip}     \fi
\ifx \shownote     \undefined \def \shownote      #1{#1}          \fi
\ifx \showarticletitle \undefined \def \showarticletitle #1{#1}   \fi
\ifx \showURL      \undefined \def \showURL       #1{#1}          \fi
\providecommand\bibfield[2]{#2}
\providecommand\bibinfo[2]{#2}
\providecommand\natexlab[1]{#1}
\providecommand\showeprint[2][]{arXiv:#2}

\bibitem[\protect\citeauthoryear{Aiello, Chung, and Lu}{Aiello
  et~al\mbox{.}}{2000}]%
        {aiello2000random}
\bibfield{author}{\bibinfo{person}{William Aiello}, \bibinfo{person}{Fan
  Chung}, {and} \bibinfo{person}{Linyuan Lu}.} \bibinfo{year}{2000}\natexlab{}.
\newblock \showarticletitle{A random graph model for massive graphs}. In
  \bibinfo{booktitle}{{\em Proceedings of the thirty-second annual ACM
  symposium on Theory of computing}}. Acm, \bibinfo{pages}{171--180}.
\newblock


\bibitem[\protect\citeauthoryear{Amanatidis, Green, and Mihail}{Amanatidis
  et~al\mbox{.}}{2015}]%
        {amanatidis2015graphic}
\bibfield{author}{\bibinfo{person}{Georgios Amanatidis},
  \bibinfo{person}{Bradley Green}, {and} \bibinfo{person}{Milena Mihail}.}
  \bibinfo{year}{2015}\natexlab{}.
\newblock \showarticletitle{Graphic realizations of joint-degree matrices}.
\newblock \bibinfo{journal}{{\em arXiv preprint arXiv:1509.07076\/}}
  (\bibinfo{year}{2015}).
\newblock


\bibitem[\protect\citeauthoryear{Bassler, Del~Genio, Erd{\H{o}}s, Mikl{\'o}s,
  and Toroczkai}{Bassler et~al\mbox{.}}{2015}]%
        {bassler2015exact}
\bibfield{author}{\bibinfo{person}{Kevin~E Bassler}, \bibinfo{person}{Charo~I
  Del~Genio}, \bibinfo{person}{P{\'e}ter~L Erd{\H{o}}s},
  \bibinfo{person}{Istv{\'a}n Mikl{\'o}s}, {and} \bibinfo{person}{Zolt{\'a}n
  Toroczkai}.} \bibinfo{year}{2015}\natexlab{}.
\newblock \showarticletitle{Exact sampling of graphs with prescribed degree
  correlations}.
\newblock \bibinfo{journal}{{\em New Journal of Physics\/}}
  \bibinfo{volume}{17}, \bibinfo{number}{8} (\bibinfo{year}{2015}),
  \bibinfo{pages}{083052}.
\newblock


\bibitem[\protect\citeauthoryear{Blitzstein and Diaconis}{Blitzstein and
  Diaconis}{2011}]%
        {blitzstein2011sequential}
\bibfield{author}{\bibinfo{person}{Joseph Blitzstein} {and}
  \bibinfo{person}{Persi Diaconis}.} \bibinfo{year}{2011}\natexlab{}.
\newblock \showarticletitle{A sequential importance sampling algorithm for
  generating random graphs with prescribed degrees}.
\newblock \bibinfo{journal}{{\em Internet Mathematics\/}} \bibinfo{volume}{6},
  \bibinfo{number}{4} (\bibinfo{year}{2011}), \bibinfo{pages}{489--522}.
\newblock


\bibitem[\protect\citeauthoryear{Czabarka, Dutle, Erd{\H{o}}s, and
  Mikl{\'o}s}{Czabarka et~al\mbox{.}}{2015}]%
        {czabarka2015realizations}
\bibfield{author}{\bibinfo{person}{{\'E}va Czabarka}, \bibinfo{person}{Aaron
  Dutle}, \bibinfo{person}{P{\'e}ter~L Erd{\H{o}}s}, {and}
  \bibinfo{person}{Istv{\'a}n Mikl{\'o}s}.} \bibinfo{year}{2015}\natexlab{}.
\newblock \showarticletitle{On realizations of a joint degree matrix}.
\newblock \bibinfo{journal}{{\em Discrete Applied Mathematics\/}}
  \bibinfo{volume}{181} (\bibinfo{year}{2015}), \bibinfo{pages}{283--288}.
\newblock


\bibitem[\protect\citeauthoryear{Del~Genio, Kim, Toroczkai, and
  Bassler}{Del~Genio et~al\mbox{.}}{2010}]%
        {del2010efficient}
\bibfield{author}{\bibinfo{person}{Charo~I Del~Genio}, \bibinfo{person}{Hyunju
  Kim}, \bibinfo{person}{Zolt{\'a}n Toroczkai}, {and} \bibinfo{person}{Kevin~E
  Bassler}.} \bibinfo{year}{2010}\natexlab{}.
\newblock \showarticletitle{Efficient and exact sampling of simple graphs with
  given arbitrary degree sequence}.
\newblock \bibinfo{journal}{{\em PloS one\/}} \bibinfo{volume}{5},
  \bibinfo{number}{4} (\bibinfo{year}{2010}), \bibinfo{pages}{e10012}.
\newblock


\bibitem[\protect\citeauthoryear{Devanny, Eppstein, and Tillman}{Devanny
  et~al\mbox{.}}{2016}]%
        {devanny2016computational}
\bibfield{author}{\bibinfo{person}{William Devanny}, \bibinfo{person}{David
  Eppstein}, {and} \bibinfo{person}{B{\'a}lint Tillman}.}
  \bibinfo{year}{2016}\natexlab{}.
\newblock \showarticletitle{The computational hardness of dK-series}. In
  \bibinfo{booktitle}{{\em NetSci 2016}}.
\newblock


\bibitem[\protect\citeauthoryear{Dimitropoulos, Krioukov, Vahdat, and
  Riley}{Dimitropoulos et~al\mbox{.}}{2009}]%
        {Dimitropoulos:2009:GAM:1596519.1596522}
\bibfield{author}{\bibinfo{person}{Xenofontas Dimitropoulos},
  \bibinfo{person}{Dmitri Krioukov}, \bibinfo{person}{Amin Vahdat}, {and}
  \bibinfo{person}{George Riley}.} \bibinfo{year}{2009}\natexlab{}.
\newblock \showarticletitle{Graph Annotations in Modeling Complex Network
  Topologies}.
\newblock \bibinfo{journal}{{\em ACM Trans. Model. Comput. Simul.\/}}
  \bibinfo{volume}{19}, \bibinfo{number}{4}, Article \bibinfo{articleno}{17}
  (\bibinfo{date}{Nov.} \bibinfo{year}{2009}), \bibinfo{numpages}{29}~pages.
\newblock
\showISSN{1049-3301}
\showDOI{%
\url{http://dx.doi.org/10.1145/1596519.1596522}}


\bibitem[\protect\citeauthoryear{Dorogovtsev}{Dorogovtsev}{2003}]%
        {dorogovtsev2003networks}
\bibfield{author}{\bibinfo{person}{SN Dorogovtsev}.}
  \bibinfo{year}{2003}\natexlab{}.
\newblock \showarticletitle{Networks with desired correlations}.
\newblock \bibinfo{journal}{{\em arXiv preprint cond-mat/0308336\/}}
  (\bibinfo{year}{2003}).
\newblock


\bibitem[\protect\citeauthoryear{Erd{\H{o}}s and Gallai}{Erd{\H{o}}s and
  Gallai}{1960}]%
        {erdHos1960grafok}
\bibfield{author}{\bibinfo{person}{P Erd{\H{o}}s} {and} \bibinfo{person}{T
  Gallai}.} \bibinfo{year}{1960}\natexlab{}.
\newblock \showarticletitle{Gr{\'a}fok el{\H{o}}{\'\i}rt fok{\'u} pontokkal}.
\newblock \bibinfo{journal}{{\em Mat. Lapok\/}}  \bibinfo{volume}{11}
  (\bibinfo{year}{1960}), \bibinfo{pages}{264--274}.
\newblock


\bibitem[\protect\citeauthoryear{Erd{\H{o}}s, Hartke, van Iersel, and
  Mikl{\'o}s}{Erd{\H{o}}s et~al\mbox{.}}{2015}]%
        {erdHos2015graph}
\bibfield{author}{\bibinfo{person}{P{\'e}ter~L Erd{\H{o}}s},
  \bibinfo{person}{Stephen~G Hartke}, \bibinfo{person}{Leo van Iersel}, {and}
  \bibinfo{person}{Istv{\'a}n Mikl{\'o}s}.} \bibinfo{year}{2015}\natexlab{}.
\newblock \showarticletitle{Graph realizations constrained by skeleton graphs}.
\newblock \bibinfo{journal}{{\em arXiv preprint arXiv:1508.00542\/}}
  (\bibinfo{year}{2015}).
\newblock


\bibitem[\protect\citeauthoryear{Erd{\H{o}}s, Kir{\'a}ly, and
  Mikl{\'o}s}{Erd{\H{o}}s et~al\mbox{.}}{2013}]%
        {erdHos2013swap}
\bibfield{author}{\bibinfo{person}{P{\'e}ter~L Erd{\H{o}}s},
  \bibinfo{person}{Zolt{\'a}n Kir{\'a}ly}, {and} \bibinfo{person}{Istv{\'a}n
  Mikl{\'o}s}.} \bibinfo{year}{2013}\natexlab{}.
\newblock \showarticletitle{On the swap-distances of different realizations of
  a graphical degree sequence}.
\newblock \bibinfo{journal}{{\em Combinatorics, Probability and Computing\/}}
  \bibinfo{volume}{22}, \bibinfo{number}{03} (\bibinfo{year}{2013}),
  \bibinfo{pages}{366--383}.
\newblock


\bibitem[\protect\citeauthoryear{Erd{\H{o}}s and Mikl{\'o}s}{Erd{\H{o}}s and
  Mikl{\'o}s}{2016}]%
        {erdHos2016not}
\bibfield{author}{\bibinfo{person}{P{\'e}ter~L Erd{\H{o}}s} {and}
  \bibinfo{person}{Istv{\'a}n Mikl{\'o}s}.} \bibinfo{year}{2016}\natexlab{}.
\newblock \showarticletitle{Not all simple looking degree sequence problems are
  easy}.
\newblock \bibinfo{journal}{{\em arXiv preprint arXiv:1606.00730\/}}
  (\bibinfo{year}{2016}).
\newblock


\bibitem[\protect\citeauthoryear{Erdos, Mikl{\'o}s, and Toroczkai}{Erdos
  et~al\mbox{.}}{2015}]%
        {erdos2015decomposition}
\bibfield{author}{\bibinfo{person}{P{\'e}ter~L Erdos},
  \bibinfo{person}{Istv{\'a}n Mikl{\'o}s}, {and} \bibinfo{person}{Zolt{\'a}n
  Toroczkai}.} \bibinfo{year}{2015}\natexlab{}.
\newblock \showarticletitle{A decomposition based proof for fast mixing of a
  Markov chain over balanced realizations of a joint degree matrix}.
\newblock \bibinfo{journal}{{\em SIAM Journal on Discrete Mathematics\/}}
  \bibinfo{volume}{29}, \bibinfo{number}{1} (\bibinfo{year}{2015}),
  \bibinfo{pages}{481--499}.
\newblock


\bibitem[\protect\citeauthoryear{Erd{\H{o}}s, Mikl{\'o}s, and
  Toroczkai}{Erd{\H{o}}s et~al\mbox{.}}{2016}]%
        {erdHos2016new}
\bibfield{author}{\bibinfo{person}{P{\'e}ter~L Erd{\H{o}}s},
  \bibinfo{person}{Istv{\'a}n Mikl{\'o}s}, {and} \bibinfo{person}{Zolt{\'a}n
  Toroczkai}.} \bibinfo{year}{2016}\natexlab{}.
\newblock \showarticletitle{New classes of degree sequences with fast mixing
  swap Markov chain sampling}.
\newblock \bibinfo{journal}{{\em arXiv preprint arXiv:1601.08224\/}}
  (\bibinfo{year}{2016}).
\newblock


\bibitem[\protect\citeauthoryear{Fulkerson et~al\mbox{.}}{Fulkerson
  et~al\mbox{.}}{1960}]%
        {fulkerson1960zero}
\bibfield{author}{\bibinfo{person}{Delbert~Ray Fulkerson} {and}
  \bibinfo{person}{others}.} \bibinfo{year}{1960}\natexlab{}.
\newblock \showarticletitle{Zero-one matrices with zero trace}.
\newblock \bibinfo{journal}{{\em Pacific J. Math\/}} \bibinfo{volume}{10},
  \bibinfo{number}{3} (\bibinfo{year}{1960}), \bibinfo{pages}{831--836}.
\newblock


\bibitem[\protect\citeauthoryear{Gale et~al\mbox{.}}{Gale
  et~al\mbox{.}}{1957}]%
        {gale1957theorem}
\bibfield{author}{\bibinfo{person}{David Gale} {and} \bibinfo{person}{others}.}
  \bibinfo{year}{1957}\natexlab{}.
\newblock \showarticletitle{A theorem on flows in networks}.
\newblock \bibinfo{journal}{{\em Pacific J. Math\/}} \bibinfo{volume}{7},
  \bibinfo{number}{2} (\bibinfo{year}{1957}), \bibinfo{pages}{1073--1082}.
\newblock


\bibitem[\protect\citeauthoryear{Gjoka, Kurant, and Markopoulou}{Gjoka
  et~al\mbox{.}}{2013}]%
        {gjoka20132}
\bibfield{author}{\bibinfo{person}{Minas Gjoka}, \bibinfo{person}{Maciej
  Kurant}, {and} \bibinfo{person}{Athina Markopoulou}.}
  \bibinfo{year}{2013}\natexlab{}.
\newblock \showarticletitle{2.5 k-graphs: from sampling to generation}. In
  \bibinfo{booktitle}{{\em INFOCOM, 2013 Proceedings IEEE}}. IEEE,
  \bibinfo{pages}{1968--1976}.
\newblock


\bibitem[\protect\citeauthoryear{Gjoka, Tillman, and Markopoulou}{Gjoka
  et~al\mbox{.}}{2015}]%
        {gjoka2015construction}
\bibfield{author}{\bibinfo{person}{Minas Gjoka}, \bibinfo{person}{B{\'a}lint
  Tillman}, {and} \bibinfo{person}{Athina Markopoulou}.}
  \bibinfo{year}{2015}\natexlab{}.
\newblock \showarticletitle{Construction of simple graphs with a target joint
  degree matrix and beyond}. In \bibinfo{booktitle}{{\em 2015 IEEE Conference
  on Computer Communications (INFOCOM)}}. IEEE, \bibinfo{pages}{1553--1561}.
\newblock


\bibitem[\protect\citeauthoryear{Hagberg, Schult, and Swart}{Hagberg
  et~al\mbox{.}}{2008}]%
        {hagberg-2008-exploring}
\bibfield{author}{\bibinfo{person}{Aric~A. Hagberg}, \bibinfo{person}{Daniel~A.
  Schult}, {and} \bibinfo{person}{Pieter~J. Swart}.}
  \bibinfo{year}{2008}\natexlab{}.
\newblock \showarticletitle{Exploring network structure, dynamics, and function
  using {NetworkX}}. In \bibinfo{booktitle}{{\em Proceedings of the 7th Python
  in Science Conference (SciPy2008)}}. \bibinfo{address}{Pasadena, CA USA},
  \bibinfo{pages}{11--15}.
\newblock


\bibitem[\protect\citeauthoryear{Hakimi}{Hakimi}{1962}]%
        {hakimi1962realizability}
\bibfield{author}{\bibinfo{person}{S~Louis Hakimi}.}
  \bibinfo{year}{1962}\natexlab{}.
\newblock \showarticletitle{On realizability of a set of integers as degrees of
  the vertices of a linear graph. I}.
\newblock \bibinfo{journal}{{\it J. Soc. Indust. Appl. Math.}}
  \bibinfo{volume}{10}, \bibinfo{number}{3} (\bibinfo{year}{1962}),
  \bibinfo{pages}{496--506}.
\newblock


\bibitem[\protect\citeauthoryear{Havel}{Havel}{1955}]%
        {havel1955poznamka}
\bibfield{author}{\bibinfo{person}{V{\'a}clav Havel}.}
  \bibinfo{year}{1955}\natexlab{}.
\newblock \showarticletitle{Pozn{\'a}mka o existenci konecnych grafu}.
\newblock \bibinfo{journal}{{\em {\v{C}}asopis pro p{\v{e}}stov{\'a}n{\'\i}
  matematiky\/}} \bibinfo{volume}{80}, \bibinfo{number}{4}
  (\bibinfo{year}{1955}), \bibinfo{pages}{477--480}.
\newblock


\bibitem[\protect\citeauthoryear{Holland and Leinhardt}{Holland and
  Leinhardt}{1976}]%
        {holland1976local}
\bibfield{author}{\bibinfo{person}{Paul~W Holland} {and}
  \bibinfo{person}{Samuel Leinhardt}.} \bibinfo{year}{1976}\natexlab{}.
\newblock \showarticletitle{Local structure in social networks}.
\newblock \bibinfo{journal}{{\em Sociological methodology\/}}
  \bibinfo{volume}{7} (\bibinfo{year}{1976}), \bibinfo{pages}{1--45}.
\newblock


\bibitem[\protect\citeauthoryear{Kim, Del~Genio, Bassler, and Toroczkai}{Kim
  et~al\mbox{.}}{2012}]%
        {kim2012constructing}
\bibfield{author}{\bibinfo{person}{Hyunju Kim}, \bibinfo{person}{Charo~I
  Del~Genio}, \bibinfo{person}{Kevin~E Bassler}, {and}
  \bibinfo{person}{Zolt{\'a}n Toroczkai}.} \bibinfo{year}{2012}\natexlab{}.
\newblock \showarticletitle{Constructing and sampling directed graphs with
  given degree sequences}.
\newblock \bibinfo{journal}{{\em New Journal of Physics\/}}
  \bibinfo{volume}{14}, \bibinfo{number}{2} (\bibinfo{year}{2012}),
  \bibinfo{pages}{023012}.
\newblock


\bibitem[\protect\citeauthoryear{Leskovec and Krevl}{Leskovec and
  Krevl}{2014}]%
        {snapnets}
\bibfield{author}{\bibinfo{person}{Jure Leskovec} {and} \bibinfo{person}{Andrej
  Krevl}.} \bibinfo{year}{2014}\natexlab{}.
\newblock \bibinfo{title}{{SNAP Datasets}: {Stanford} Large Network Dataset
  Collection}.
\newblock \bibinfo{howpublished}{\url{http://snap.stanford.edu/data}}.
  (\bibinfo{date}{June} \bibinfo{year}{2014}).
\newblock


\bibitem[\protect\citeauthoryear{Mahadevan, Krioukov, Fall, and
  Vahdat}{Mahadevan et~al\mbox{.}}{2006}]%
        {mahadevan2006systematic}
\bibfield{author}{\bibinfo{person}{Priya Mahadevan}, \bibinfo{person}{Dmitri
  Krioukov}, \bibinfo{person}{Kevin Fall}, {and} \bibinfo{person}{Amin
  Vahdat}.} \bibinfo{year}{2006}\natexlab{}.
\newblock \showarticletitle{Systematic topology analysis and generation using
  degree correlations}. In \bibinfo{booktitle}{{\em ACM SIGCOMM Computer
  Communication Review}}, Vol.~\bibinfo{volume}{36}. ACM,
  \bibinfo{pages}{135--146}.
\newblock


\bibitem[\protect\citeauthoryear{Orsini, Dankulov, Colomer-de Sim{\'o}n,
  Jamakovic, Mahadevan, Vahdat, Bassler, Toroczkai, Bogu{\~n}{\'a}, Caldarelli,
  et~al\mbox{.}}{Orsini et~al\mbox{.}}{2015}]%
        {orsini2015quantifying}
\bibfield{author}{\bibinfo{person}{Chiara Orsini}, \bibinfo{person}{Marija~M
  Dankulov}, \bibinfo{person}{Pol Colomer-de Sim{\'o}n},
  \bibinfo{person}{Almerima Jamakovic}, \bibinfo{person}{Priya Mahadevan},
  \bibinfo{person}{Amin Vahdat}, \bibinfo{person}{Kevin~E Bassler},
  \bibinfo{person}{Zolt{\'a}n Toroczkai}, \bibinfo{person}{Mari{\'a}n
  Bogu{\~n}{\'a}}, \bibinfo{person}{Guido Caldarelli}, {and}
  \bibinfo{person}{others}.} \bibinfo{year}{2015}\natexlab{}.
\newblock \showarticletitle{Quantifying randomness in real networks}.
\newblock \bibinfo{journal}{{\em Nature communications\/}}  \bibinfo{volume}{6}
  (\bibinfo{year}{2015}).
\newblock


\bibitem[\protect\citeauthoryear{Snijders, Pattison, Robins, and
  Handcock}{Snijders et~al\mbox{.}}{2006}]%
        {snijders.et.al:sm:2006}
\bibfield{author}{\bibinfo{person}{Tom A.~B. Snijders},
  \bibinfo{person}{Philippa~E. Pattison}, \bibinfo{person}{Garry~L. Robins},
  {and} \bibinfo{person}{Mark~S. Handcock}.} \bibinfo{year}{2006}\natexlab{}.
\newblock \showarticletitle{New Specifications for Exponential Random Graph
  Models}.
\newblock \bibinfo{journal}{{\em Sociological Methodology\/}}
  \bibinfo{volume}{36} (\bibinfo{year}{2006}), \bibinfo{pages}{99{--}154}.
\newblock


\bibitem[\protect\citeauthoryear{Stanton and Pinar}{Stanton and Pinar}{2012}]%
        {stanton2012constructing}
\bibfield{author}{\bibinfo{person}{Isabelle Stanton} {and} \bibinfo{person}{Ali
  Pinar}.} \bibinfo{year}{2012}\natexlab{}.
\newblock \showarticletitle{Constructing and sampling graphs with a prescribed
  joint degree distribution}.
\newblock \bibinfo{journal}{{\em Journal of Experimental Algorithmics (JEA)\/}}
   \bibinfo{volume}{17} (\bibinfo{year}{2012}), \bibinfo{pages}{3--5}.
\newblock


\bibitem[\protect\citeauthoryear{Taylor}{Taylor}{1980}]%
        {taylor1980constrained}
\bibfield{author}{\bibinfo{person}{R Taylor}.} \bibinfo{year}{1980}\natexlab{}.
\newblock \bibinfo{booktitle}{{\em Constrained switchings in graphs}}.
\newblock \bibinfo{publisher}{University of Melbourne, Department of
  Mathematics}.
\newblock


\end{thebibliography}

\end{document}